# Identification of Mars gully activity types associated with ice composition.

*Mathieu Vincendon*

Institut d'Astrophysique Spatiale, Université Paris-Sud, Orsay, France (mathieu.vincendon@u-psud.fr)

## Key points

- We characterize the presence and composition of ice at currently active gullies.
- Major channel incisions systematically occur where $CO_2$ ice is observed or probable.
- Some activity (e.g. bright deposits) is not associated with $CO_2$ but with $H_2O$ ice.

## Abstract

The detection of geologically recent channels at the end of the twentieth century rapidly suggested that liquid water could have been present on Mars up to recent times. A mechanism involving melting of water ice during ice ages in the last several million years progressively emerged during years following the first observations of these gullies. However, the recent discovery of current activity within gullies now suggests a paradigm shift where a contemporary $CO_2$ ice-based and liquid water-free mechanism may form all gullies. Here we perform a survey of near-infrared observations and construct time sequences of water and $CO_2$ ice formation and sublimation at active gully sites. We observe that all major new erosive features such as channel development or lengthening systematically occur where and, if applicable, when $CO_2$ ice is observed or probable. $CO_2$ ice layers are however estimated to be only 1 mm to 1 cm thick for low-latitude sites, which may have implication for potential formation mechanisms. We also observe that part of current gully activity, notably the formation of some new deposits, is poorly compatible with the presence of $CO_2$ ice. In particular, all new bright deposits reported in the literature have a low $CO_2$ ice probability while water ice should be present at most sites. Our results confirm that $CO_2$ ice is a key factor controlling present-day channel development on Mars and show that other mechanisms, potentially involving sublimation or melting of water ice, are also contributing to current gully activity.

## 1 Introduction

The discovery of numerous geologically recent downslope channels in images with scales of meters per pixel acquired at the end of the twentieth century aroused strong interest among the Martian scientific community (Malin & Edgett, 2000). A large number of possible formation mechanisms arose rapidly to explain these kilometer-scale features usually referred as "gullies." They can be classified according to the composition of the triggering factor: processes somehow involving water, liquid or solid (Malin & Edgett, 2000; Costard, et al., 2002; Hecht, 2002; Reiss & Jaumann, 2003; Christensen, 2003; Hugenholtz, 2008), $CO_2$-based mechanisms (Musselwhite, et al., 2001; Hoffman, 2002; Ishii & Sasaki, 2004; Hugenholtz, 2008; Cedillo-Flores, et al., 2011; Diniega, et al., 2013; Pilorget & Forget, 2015), and marginal formation pathways without volatiles (Treiman, 2003; Shinbrot, et al., 2004; Bart, 2007).

Gully formation time is imprecisely constrained but is securely known to be younger than a few million years for most channels (Reiss & Jaumann, 2003; Schon, et al., 2009). Modern Mars pressure and temperature ranges place the planet at the frontier between liquid water stability and instability (Haberle, et al., 2001; Kossacki & Markiewicz, 2004; Richardson & Mischna, 2005; Mohlmann, 2010). The survival of liquid water events up to recent times could have implications for the habitability of Mars: the possibility that liquid water carved these channels has thus been the subject of a wide range of scientific investigations. Some liquid-water-based formation mechanisms have been favored according to recent research (Dickson, et al., 2007; Head, et al., 2008; Williams, et al., 2009): they rely on water ice melting after geologically recent, but million years old, ice ages (Laskar & Robutel, 1993; Mellon & Jakosky, 1995; Head, et al., 2003). While these liquid-based formation mechanisms are compelling, notably in terms of channel morphology (Mangold, et al., 2003; Védie, et al., 2008; Lanza, et al., 2010; Levy, et al., 2010; Johnsson, et al., 2014), the alternative mechanisms based on granular or gas-lubricated flow within $CO_2$ ice context also proved to be satisfying explanations to several of the observed gully characteristics. Actually, recent studies have highlighted the fact that gullies form by multiple episode of activity, that several formation mechanisms are probably acting and that part of this diversity may be apparent in gully morphological differences (Hugenholtz, 2008; Schon, et al., 2009; Bryson, et al., 2010; Mangold, et al., 2010; Hobbs, et al., 2014; Conway, et al., 2015). The question of the relative contribution of the different mechanisms remains open.

Current modification at gully sites has been first reported with the observation of new bright deposits within pre-existing mid-latitude gully channels (Malin, et al., 2006). The High Resolution Imaging Science Experiment (HiRISE) (McEwen, et al., 2007a) has monitored known gully sites since arrival in orbit in 2006. Analysis of observations revealed ongoing modification with various modification types: new bright, dark, colorful or neutral deposits, some of which are topographically resolved; erosion in the form of increases of channel and alcove length and width; and a few indisputable occurrences of new gullies or channels (Harrison, et al., 2009; Reiss, et al., 2010; Dundas, et al., 2010; Diniega, et al., 2010; Hansen, et al., 2011; Dundas, et al., 2012; Raack, et al., 2015; Dundas, et al., 2015). This discovery highlighted the fact that conditions for the development of preexisting gullies systems are met today (Dundas, et al., 2015). Two possibilities are thus debated: either current activity is the main gully formation mechanism, able to create highly complex, large and mature channel systems over long periods, or we are observing a mechanism that may only modify otherwise formed gully systems (Dickson, et al., 2015; Dundas, et al., 2015). The first scenario would imply that gullies observed today did not mainly form during past glaciations through water ice melting.

Gullies are preferentially found where ice is more likely to form either at present or with a different past obliquity: gullies are almost never observed equatorward 30° latitude and essentially occur on pole-facing slopes for latitudes lower than 45° (Heldmann & Mellon, 2004; Harrison, et al., 2015). This characteristic is one of the key arguments in favor of a volatile-based main formation mechanism. A few gully-like landforms have also been reported near the equator (Treiman, 2003; Dundas, et al., 2015) but they are marginal and ice is anyhow predicted at equatorial latitudes at high obliquity (Forget, et al., 2006). The observed spatial distribution of seasonal ice and frost matches at global scale the distribution of most known gullies, including active gullies (Schorghofer & Edgett, 2006; Vincendon, et al., 2010a; Vincendon, et al., 2010b). Gully changes for which time of event is constrained occur mainly during winter (Diniega, et al., 2010; Dundas, et al., 2010; Dundas, et al., 2012). Seasonal ice is indeed frequently directly observed where and when major gully modifications occur (Reiss, et al., 2010; Hansen, et al., 2011; Dundas, et al., 2012; Dundas, et al., 2015). As seasonal ice is mostly composed of $CO_2$, it has been argued that $CO_2$ ice should be the main responsible agent for current gully activity (Dundas, et al., 2012; Diniega, et al., 2013). This is notably supported by the observation of ephemeral dark spots and dark flows over seasonal frost at some active gully sites (Gardin, et al., 2010; Dundas, et al., 2012; Jouannic, et al., 2012; Raack, et al., 2015; Hansen, et al., 2011). These features, common at higher latitudes (Kieffer, 2007), are indicative of the presence of $CO_2$ jets that are part of some recent $CO_2$ formation mechanisms for gullies (Pilorget & Forget, 2015).

While $CO_2$ is indeed the main component of seasonal frost, ice layers uniquely made of water can occur at latitudes of gully activity, notably during spring and at low latitudes (Vincendon, et al., 2010a; Vincendon, et al., 2010b). Preliminary analyses of ice composition at a few active gullies indeed revealed discrepancies between location/timing of reported activity and presence of $CO_2$ ice (Vincendon, et al., 2013; Vincendon, et al., 2014). In addition to the possibility that a few recent gully changes might be due to non-icy processes (Dundas, et al., 2010), it has been suggested that incomplete seasonal coverage, or insufficient spatial sampling, might be the reason for the nonobservation of $CO_2$ ice at some active gully sites (Dundas, et al., 2015). In this paper, we aim at characterizing the timing of water and $CO_2$ ice condensation and sublimation at previously reported active gully sites. We first perform an exhaustive survey of available near-infrared imaging spectroscopy data. We then compare, as a function of time during a year, available constraints about gully activity and ice formation and sublimation, to determine if ice (and which ice) is present during activity. A special attention is devoted to low-latitude sites, where $CO_2$ ice has not yet been reported or is not expected. We also analyze in more detail sites for which precise time constraints have been obtained, so as to perform relevant comparisons between activity and ice timings.

## 2 Observations

### 2.1 Method

Our study is dedicated to southern active gullies located equatorward 50°S, a limit that roughly separate polar and nonpolar gullies (Harrison, et al., 2015). The number of detected active gullies (as well as gullies in general) drops poleward approximately 55°, with missing information about the timing of changes between 50° and 55° (Dundas, et al., 2015) except for the Russell crater dunes for which dedicated studies have been conducted (Gardin, et al., 2010; Reiss, et al., 2010; Jouannic, et al., 2012). This lack of timing constraints also concerns the equatorward northern site at 51.7°N (Dundas, et al., 2015). We first drew up an exhaustive list of active gully sites based on previous studies. We extracted from the literature information concerning the location, timing, and type of modifications. This list is summarized in Table I.

Table I: List and properties of gully sites with current activity identified by other studies (left eight columns) and ice probability levels derived in this study (two right columns). Used publications for left columns are: Dundas et al. 2010/2012/2015 (D10/D12/D15); Diniega et al. 2010 (Di10); Malin et al. 2006 (Ma06); McEwen et al. 2007b (McE07) (additional personal communications by C. Dundas are also indicated by "Dpc").

| Name | Latitude | Longitude | Deposit, flow | Erosion | Change Ls range | Facing | Publications | H2O | CO2 |
|---|---|---|---|---|---|---|---|---|---|
| Kibuye crater | 29.1°S | 181.8°E | Dark | | 18-288 and 300-100 | S | D10, D12, D15 | 5 | 1 |
| Promethei Terra | 32.3°S | 118.6°E | Dark | | | S | D12, D15 | 5 | 5 |
| Noachis Terra #1 | 32.4°S | 338.2°E | Color | | | S, SW | D12, D15 | 2 | 2 |
| Kandi crater | 32.7°S | 122.1°E | Deposit | | 10-200 | SE | D15 | 5 | 3 |
| Ariadnes Colles | 34.4°S | 172.3°E | Neutral (dark?) | | 149-97 and 60-110 | S | D10, D12, D15 | 3 | 5 |
| Gasa crater | 35.7°S | 129.4°E | Dark | | 65-109 and 109-152 | S | D10, D12, D15 | 5 | 5 |
| | | | Color | | 65-152 | | | 5 | 5 |
| | | | Bright | | 152-169 | | | 5 | 1 |
| | | | Deposit | | 130-140 | | | 5 | 5 |
| | | | | Channel widening | 109-152 | | | 5 | 5 |
| | | | | Slump. alcove material | 212-312 | | | 1 | 1 |
| Noachis Terra #2 | 35.8°S | 330.8°E | Bright | | | E | D12, D15 | 3 | 3 |
| Terra Sirenum #1 | 36.0°S | 214.3°E | Deposit | | 0-320 | S | D15 | 5 | 3 |
| Naruko crater | 36.2°S | 198.3°E | Bright | | | SE | Ma06 | 5 | 3 |
| Near Gorgonum #1 | 36.4°S | 190.4°E | Dark | | 328-234 | SE | D10 | 5 | 4 |
| Gorgonum Chaos | 37.2°S | 188.3°E | Neutral (dark?) | | 18-150 and 60-100 | SE, E | D12, D15 | 3 | 3 |
| Pursat crater | 37.4°S | 130.7°E | Bright | | 28-209 | NE | D10, D12 | 3 | 3 |
| Terra Sirenum #2 | 37.4°S | 229.0°E | Neutral / Thick | | 5-120 | S-SW | D15 | 5 | 3 |
| | | | Bright | Fresh looking | | SW-W | McE07 | 4 | 2 |
| Terra Sirenum #3 | 37.5°S | 222.9°E | Thick flow | Major channel incision | | S | D15 | 5 | 4 |
| Near Gorgonum #2 | 37.6°S | 192.9°E | Thick flow | Channel incision | | S | D15 | 5 | 5 |
| Terra Sirenum #4 | 38.1°S | 215.9°E | | Channel incision | | SW | D15 | 5 | 5 |
| Terra Sirenum #5 | 38.1°S | 224°E | Thick flow | Channel incision | 0-230 | SE, SW | D15 | 5 | 4 |
| Penticton crater | 38.4°S | 96.8°E | Bright | | | NW | Ma06 | 5 | 1 |
| Simois Colles | 38.6°S | 183.8°E | Deposit | | | SE | D15 | 5 | 5 |
| Niquero crater | 38.7°S | 194.0°E | Deposit | | | SE | D15 | 5 | 5 |
| Corozal crater | 38.8°S | 159.5°E | Dark | | 354-312 (61-183?) | S-SW | D10, D15 | 5 | 5 |
| | | | | Channel incision | | | | 5 | 5 |
| Terra Sirenum #6 | 38.9°S | 223.7°E | | Channel incision | 136-158 (44-158?) | SW | D12, Dpc | 5 | 3 |
| | | | Dark | | 136-158 | | | 5 | 1 |
| Arrhenius crater | 39.6°S | 123.1°E | Deposit | | 100-180 | SW | D15 | 5 | 5 |
| Terra Sirenum #7 | 40.3°S | 217.1°E | Topo change | | | SW | D15 | 5 | 5 |
| Argyre | 40.5°S | 309.9°E | | | | S-SE ? | D15 | 4 | 4 |
| Avire crater | 40.8°S | 200.3°E | | New linear channels | | SW | D12 | 5 | 5 |
| Near Avire #1 | 41.1°S | 203.5°E | | | | SW | D15, Dpc | 5 | 5 |
| Near Avire #2 | 41.1°S | 189.0°E | | Channel incision | | SW | D15 | 5 | 5 |
| Palikir crater | 41.4°S | 202.3°E | Deposit | Channel widening | 340-190 | SW | D12 | 5 | 5 |
| Roseau crater | 41.7°S | 150.6°E | Bright | | 354-167 | E | McE07, D10 | 3 | 3 |
| Near Proctor #1 | 45.8°S | 36.7°E | | | | S ? | D15 | 4 | 4 |
| Kaiser crater | 46.7°S | 20.1°E | Dark (over ice?) | | 324-60 and 60-62 | W | Di10, D12 | 1 | 1 |
| | | | Dark (over ice?) | | 62-87 and 87-127 | | | 1 | 5 |
| | | | | New channels | 119-127 | | | 1 | 5 |
| | | | | Channel & alcove exp. | 146-163 | | | 5 | 5 |
| | | | | Slumping material | 62-87 and 104-119 | | | 1 | 5 |
| Asimov crater | 46.9°S | 4.3°E | Dark | | | NW | D15 | 3 | 3 |
| | | | Bright | Fresh looking | | NW | McE07 | 3 | 3 |
| Near Proctor #2 | 47.2°S | 34.0°E | | | | | D15 | 5 | 5 |
| Proctor crater | 47.5°S | 30.4°E | | New channel | | SW-W | Di10 | 5 | 5 |
| West Argyre | 48°S | 303.7°E | | | | W-SW | D15, Dpc | 5 | 5 |
| Near Proctor #3 | 49.0°S | 27.2°E | | Channel & alcove exp. | 152-179 | S | D12 | 5 | 5 |
| | | | | Gully end/apron | 179-195 | | | 5 | 5 |
| | | | | Apron changes | >195 | | | 2 | 2 |
| Matara crater | 49.5°S | 34.8°E | Thick deposits | New alcove & channel | 136-167 | SW | Di10, D12 | 5 | 5 |
| | | | | New channel / alcove | 56-199 & 167-176 | SW | Di10, D12 | 5 | 5 |
| | | | | Apron changes | 165-183 | SW | Di10, D12 | 5 | 5 |
| | | | | Alcove erosion | 283-134-165-183 | NE | D15 | 3 | 3 |

For each of these sites, we look for all available CRISM (Compact Reconnaissance Imaging Spectrometer for Mars) (Murchie, et al., 2007) and OMEGA (Observatoire pour la Minéralogie, l'Eau, les Glaces, et l'Activité) (Bibring, et al., 2005) near-IR data that exactly cover the site of gully activity ("onsite" data). We also look for data gathered in a ± 1° latitude/longitude window to increase time coverage ("offsite" data). Water and $CO_2$ ices create spectral signatures at near-infrared wavelengths that enable identification and characterization of thin ice deposits (typically a few micrometers thick for water ice and a few hundreds of micrometers of $CO_2$ ice). Near-IR spectroscopy can detect ice on Mars even if transparent or dusty, contrary to visible imagery for which ice detection relies on brightness contrast (Bibring, et al., 2004; Langevin, et al., 2007; Appéré, et al., 2011). We look at each OMEGA and CRISM observation to assess whether and which ice is, or is not, present in the spectral data using similar detection techniques as previously described (Vincendon, et al., 2010a; Vincendon, et al., 2010b). We implement various tests on each observation to optimize detection capabilities: spatial averaging over regions of interest, spectral ratio, and identification of multiple spectral features.

The spatial sampling of CRISM data is 20 m/pix for "FRT," 40 m/pix for "HRS" and "HRL," 100 m/pix for "MSW," and 200 m/pix for "MSP" and "HSP"; it is between 300 m/pix and several km/pix for OMEGA. In most case, the spatial sampling of near-IR data will thus remain lower than the size of changes detected with high-resolution (30 cm/pix) HiRISE visible imagery. In our analysis, the terms "active gully site" and onsite will thus refer to the slope where change did occur and not necessary to the exact pixel or few pixels covering activity. When onsite data are missing, we look for ice on equivalent sites (e.g., a slope with a similar angle, thermal properties and orientation) in the 1°x1° surroundings, an area over which the longitudinal variability in ice condensation is expected to be low (Vincendon, et al., 2010a; Vincendon, et al., 2010b). Local, subpixel topography that can act as cold trap (Svitek & Murray, 1990) is inherent to gullied slopes: this fact must be accounted for in our study as the dominant, pixel-size slope of a gully site may not reflect the condition within these cold traps.

Overall, the relevance of detection or nondetection thus depends on the adequacy of site characteristics (spatial extent, small-scaled topography, change timing constraints, etc.) and available near-IR data properties (spatial sampling, time sampling, signal to noise ratio, etc.). We thus conduct for each site a discussion about this relevance, and empirically estimate a subjective probability level that water or $CO_2$ ice was present where change occurred. A qualitative scale with five levels of probability is defined, from "1" which means that we have gathered sufficient near-IR constraints pointing toward a lack of ice to "5" meaning that ice should be present onsite. In between, "3" means we cannot decide and "2"/"4" are used when we lean toward absence/presence of ice. As change timing constraints are frequently weak, the lack of timing precision is not included in the probability estimate: a change without any timing constraint will be classified as 5 if ice has been observed onsite. On the other hand, we update our probability that ice was present where change occurred if a relevant timing constraint is known. Slope orientation is an important consideration in our probability estimate: if we observe ice on the pole-facing slope, while the change occurred on a slope with a different orientation, then we assign a probability rank that is not only not 5 but also not necessarily 1 as ice could form on subpixel pole-facing topography (gully walls, ripples, etc.) as frequently observed on high-resolution visible imagery (e.g., in Matara crater) (see (Dundas, et al., 2015)). We typically reduce the probability level by 1 every 45° azimuth angle difference. This rule is then tempered by additional considerations: obvious presence/absence of small-scaled topography or subpixel ice in visible imagery, availability of high-resolution observations, clues from similar nearby active sites, etc.

## 2.2 Results

We analyze in this section available near-IR data for each of the 38 reported sites (Table I), in a sequence going essentially from equatorward to poleward latitudes. The resulting ice probability levels are indicated in Table I. Nonspecific references to "ice" in this section are referring to both $CO_2$ and $H_2O$ ices.

### 2.2.1 Kibuye crater (29.1°S)

Kibuye crater is the most equatorward site (29.1°S) for which gully activity has been reported so far (Dundas, et al., 2010). Activity has been observed twice, in MY29 somewhere between solar longitude ($L_S$) 18° and 288° (dark deposit) and again in MY30/31 between $L_S$ 300° and $L_S$ 100°. Visible images show "traces of seasonal frost" between $L_S$ 100° and 112° (Dundas, et al., 2015). Kibuye crater has been observed at high resolution (20 m/pix) at $L_S$ 99.8° (CRISM FRT 247C3). We detect localized water ice deposits consistent with frost areas observed with visible imagery. Ice is not observed offsite at $L_S$ 79.2° and onsite at $L_S$ 117.7° (CRISM MSP 2365F and C325). $CO_2$ ice is not detected despite favorable observation conditions (spatial sampling and solar longitude). Hence, we assign a probability rank of 1 for $CO_2$ ice and 5 for water ice to Kibuye crater.

### 2.2.2 Low latitude sites (32.5°S)

The next three active gully sites as we move poleward have been identified about 32.5° on south or southeast/southwest facing crater walls. The timing of change is poorly constrained. Changes are limited to new deposits.

A new dark deposit was observed east of Hellas at site "Promethei Terra" (32.3°S, 118.6°E). No ice is observed at $L_S$ 29° (OMEGA 0368_1, 700 m/pix). Water ice is then detected at $L_S$ 64° (CRISM FRT 17372). The next onsite observations obtained in 2012 at $L_S$ 89.4° and 101.7° (CRISM FRT 240D7 and 2491F) reveal for the first time $CO_2$ ice equatorward 34°S, at 32.3°S (Figure 1). While most of the pole-facing slope, notably the west part, is covered by water ice only, clear spectral evidence for $CO_2$ ice over a few kilometers is also observed at the longitude of the active gully. Water ice is then observed again alone (Figure 1) afterward at $L_S$ 141.3° and 143.8° (CRISM MSP 32AF and 3397). Ice is not detected anymore onsite at $L_S$ 174.9° (CRISM HSP 26AB6) and may have already sublimated by $L_S$ 157.5° (offsite CRISM MSP 39CC).

Site "Noachis Terra #1" (32.4°S, 338.2°E) is located at the frontier of an arid longitude range where ice has not yet been observed nor predicted at the surface (Vincendon, et al., 2010a; Vincendon, et al., 2010b). No water or $CO_2$ ice is indeed detected in available onsite (CRISM MSP 1AA66 and OMEGA 1293_2) or offsite (CRISM HSP 2474E, MSP B717, HSP 1940C, and MSP 26172) observations obtained between $L_S$ 96.5° and 146.9°. We attribute a probability level of 2, and not 1, for both ices because the available spatial sampling is limited to 200 m/pix at best with no observations obtained between $L_S$ 102.6° and 134.5°.

"Kandi crater" (32.7°S, 122.1°E) is only 200 km away from site Promethei Terra where $CO_2$ ice has been detected. Onsite, we do not detect ice at $L_S$ 59.5° (CRISM HSP 16FD7) and we observe water ice only at $L_S$ 103.4° (CRISM HSP 1946C). Offsite, we also observe only water ice signatures at $L_S$ 80.7°, 133.9°, and 137.8° (OMEGA 3237_4 and 1200_4, CRISM MSP 263D5), and then no more ice at $L_S$ 146° (CRISM HSP 2681A). The intermediate spatial sampling of available data (200m to 1 km/pixel while Kandi crater diameter if about 7 km) may yet prevent $CO_2$ ice detection. However, the activity at Kandi crater occurs on the southeast facing crater wall, while observed $CO_2$ ice at Promethei Terra was restricted the pole-facing slope. Kandi crater is thus categorized 5 for $H_2O$ while only 3 for $CO_2$.

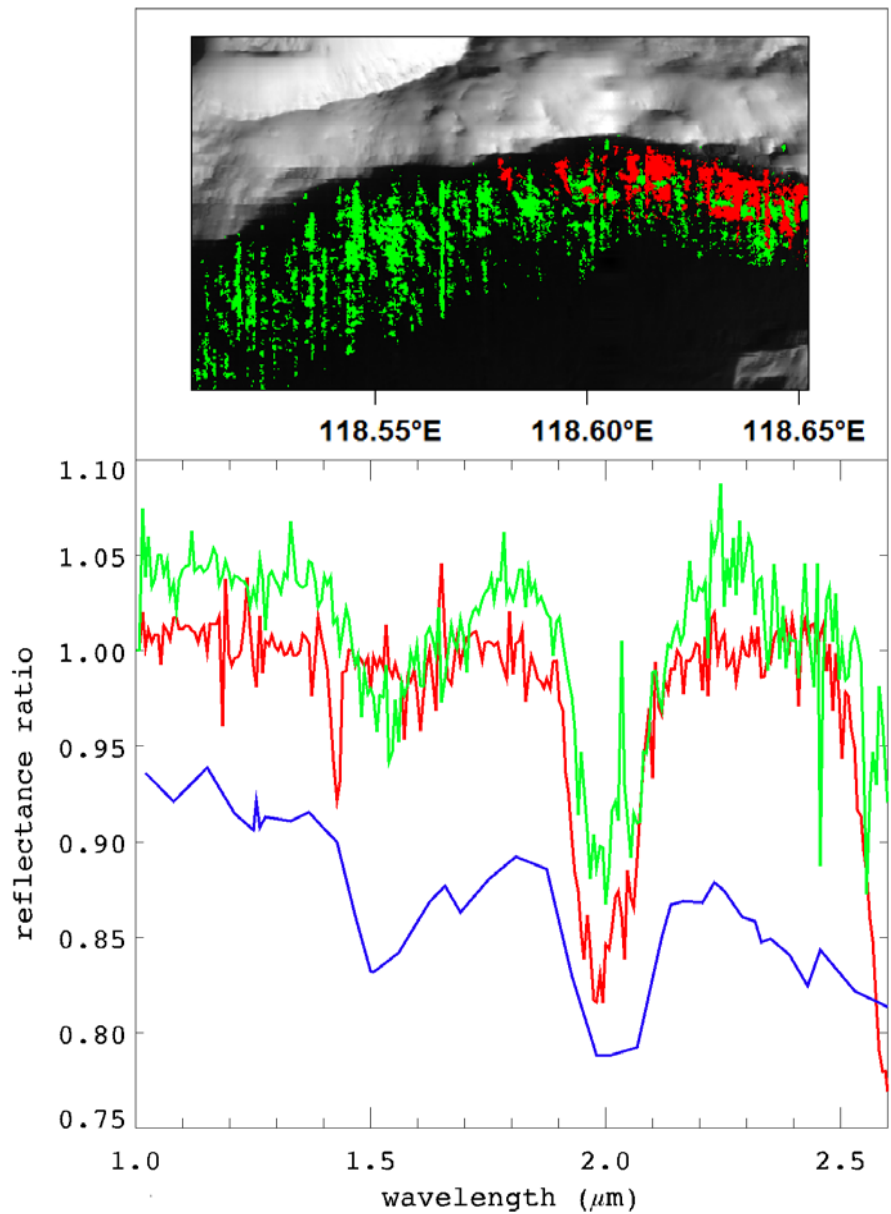

Figure 1: Ice identification on the pole-facing crater wall of site "Promethei Terra" (32.3°S). (top) Distribution of water ice (green) and $CO_2$ ice (red) at $L_S$ 101.7° (CRISM FRT 2491F). Pixels with water ice 1.5 µm and $CO_2$ ice 1.43 µm band depths (Pelkey, et al., 2007) greater than 4% and 7%, respectively, are colored. Artifacts are partly removed. North is on top. (bottom) Reflectance ratio between icy and ice-free areas. The green and red spectra are extracted from the corresponding Figure 1 (top) colored areas ($L_S$ 101.7°). The blue spectrum obtained at $L_S$ 141.3° shows only water ice signatures (CRISM MSP 32AF).

### 2.2.3 Ariadnes Colles (34.3°S)

Dark deposits have been observed twice on the pole-facing slope of a gullied mesa in Ariadnes Colles (34.3°S, 172.3°E) (Dundas, et al., 2010). Changes were first suspected to be potentially transient (Dundas, et al., 2012) but later on interpreted to be rather neutral tone and perennial (Dundas, et al., 2015). The first change (MY28/29) occurred either before $L_s$ 97° or after $L_s$ 149°; the second change (MY31) took place between $L_s$ 60° and 110°. The pole-facing slope where active gullies are observed is only 3 km wide. Ice deposits observed with visible imagery at $L_s$ 138° and 149° are only a few tens of meters wide and are restricted to the upper part of the slope (Dundas, et al., 2012). The best available spatial sampling during fall and winter is 200 m/pixel: expected deposits are thus subpixel and located in a few pixels only.

Onsite data at $L_s$ 83.2°, 96.7°, 104°, and 142.8° (respectively CRISM HSP 2DAD0, MSP B743, HSP 24ACA and 266CC) do not reveal $CO_2$ or water ice features above noise limit (typically 5–10%). At $L_s$ 137.8° (CRISM MSP CD6D), a few adjacent pixels in the upper east side of the pole-facing slope show higher values of the 1.43 μm band depth. A careful averaging of four of these pixels selected for their absence of spectral spikes reveals two major spectral features of $CO_2$ ice at 1.43 μm and 3.32 μm (9% and 40% band depth respectively) and an additional 2% absorption at 2.28 μm (Figure 2). Relative intensities are compatible with synthetic spectra of thin or small-grained $CO_2$ ice (Appéré, et al., 2011). Although water ice is not detected with a 5–10% noise limit, the longitude of Ariadnes is known for limited but probable water ice condensation on steep slopes (Vincendon, et al., 2010b). We thus assign ice probability levels of 5 ($CO_2$) and 3 (water) to Ariadnes Colles.

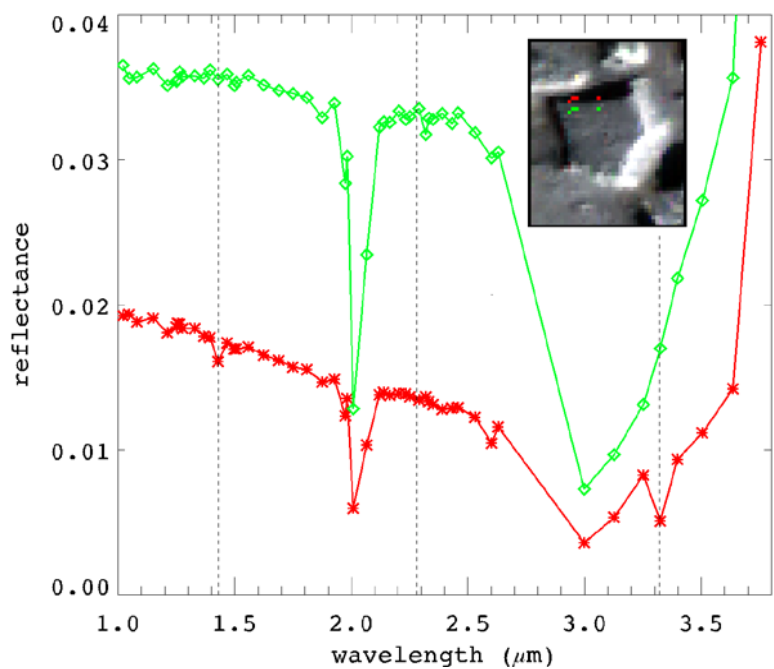

Figure 2: Reflectance spectra at Ariadnes Colles (34.3°S, 172.3°E), from observation CRISM MSP CD6D ($L_s$ 137.8°). Red: averaging of 4 pixels located on the steep upper part of the pole-facing slope. Pixels were selected for their lack of spikes among a few pixels showing higher values of the 1.43 μm band depth. Green: comparative flat surface spectra. The positions of the two main $CO_2$ ice features at 1.43 μm and 3.32 μm, plus a minor one at 2.28 μm, are indicated with dotted lines (see text for discussion). The 8 pixels selected to calculate the two mean spectra are located on the insert map (south is on top). These two reflectance "raw" spectra include atmospheric transmission (e.g., the $CO_2$ gas absorption about 2 μm).

### 2.2.4 Gasa crater (35.7°S)

Numerous types of changes have been observed at Gasa crater (deposits with various tone and erosion in channel and alcove) with precise constraints about the timing of activity (see Table I). Most of the main changes occurred between $L_s$ 65° and 169°, with one change between $L_s$ 212° and 312°.

Ice has been observed several times in visible images, from $L_S$ 109° to 152° (Dundas, et al., 2010; Dundas, et al., 2012; Dundas, et al., 2015). Gasa crater is indeed located in an area of known water ice formation (Vincendon, et al., 2010b). Near-IR observations of $CO_2$ ice have not yet been reported at this precise location, but $CO_2$ is expected at this latitude (Vincendon, et al., 2010b; Nunez, et al., 2014).

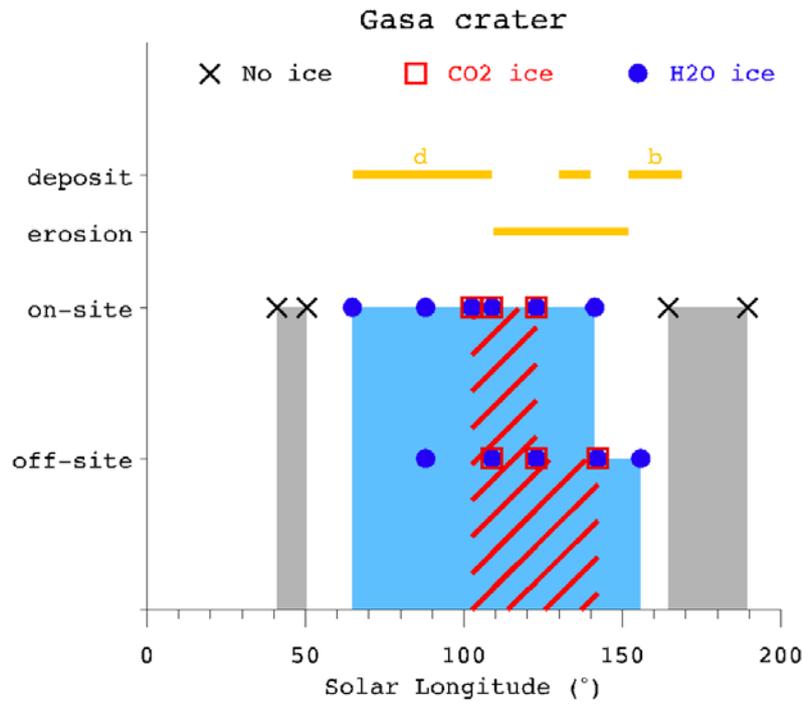

**Figure 3: Ice observations for Gasa crater pole-facing slope (35.7°S, 129.4°E) compared to change timing. Symbols indicate relevant onsite and offsite observational evidence for water ice (blue dots), $CO_2$ ice (red square), or lack of ice (grey cross). See text for details. Icy $L_S$ ranges implied by these observations are filled with the corresponding color. Timing of change (deposits or erosion) as derived from previous studies (see details in Table I) is indicated with yellow lines for comparison; letters are for deposit color ("d" for dark and "b" for bright). The slight alcove change reported between $L_S$ 212° and 312° is outside this diagram.**

Ice is not detected onsite at $L_S$ 41.1° and 50.5° (CRISM MSP A457 and FRT 1675A). Water ice is then observed alone at $L_S$ 64.9° and 88° (CRISM FRT 1742D and HSP 23FBF). $CO_2$ ice is first observed onsite, with water ice, at $L_S$ 102.6° (CRISM FRT 249AC), and then again at $L_S$ 109° and 123° (CRISM MSP 1978F and C5AE). $CO_2$ ice contaminated with water ice is restricted to a thin band in the upper part of the slope, while water ice detections are slightly more diffuse over the slope. A noisy onsite observation obtained at $L_S$ 141.4° (CRISM HSP 265F1) reveals only significant water ice signatures. However, an offsite observation (CRISM MSP 330D) still shows the presence of a few pixels with $CO_2$ ice at a similar $L_S$ (142.3°) and equatorward (35.3°S), which suggest that $CO_2$ ice may still be present on the pole-facing slope of Gasa crater at that time while being close to final sublimation. The next observation (CRISM MSP 38C8) is obtained offsite, 13.6° of $L_S$ later, and shows only water ice. $CO_2$ ice is indeed highly improbable at that time (Vincendon, et al., 2010b). No ice is anymore observed onsite at $L_S$ 164.6° (CRISM MSP 3D7B) and afterward. A summary of these observations is provided in Figure 3 and compared with change timing. The channel widening strongly correlates with the presence of $CO_2$ and water ice. The slight alcove change took place long after ice sublimation ($L_S$ > 212°). The formation of deposits correlates with the presence of at least one ice type. The timing of the bright deposits ($L_S$ 152°–169°) is not compatible with the presence of $CO_2$ ice while it corresponds to the range over which water ice is present and then sublimate.

### 2.2.5 New deposits at 36°S–37°S

Seven sites with new deposits have been reported between 35.8°S and 37.4°S. Most of these sites are characterized by nonexact pole-facing orientation and no or weak constraints about change seasonality for five of the seven sites.

The new bright deposit reported for site "Noachis Terra #2" (35.8°S, 330.8°E) is located on an east facing crater wall. The only exploitable onsite observation (CRISM HSP 2665C, $L_S$ 142.1°) reveals signatures of water and $CO_2$ ice restricted to the pole-facing crater wall, which leads us to assign ice probability levels of 3 to account for the slope orientation difference (see section 2.1).

The new deposit of site "Terra Sirenum #1" (36°S, 214.3°E) is located on the pole-facing wall of a crater. Onsite, we do not observe ice at $L_S$ 48.1°, we detect water ice only at $L_S$ 84.5°, 140.5° and 146° and no more ice at $L_S$ 169.7° (CRISM FRT 16570, HSP 2DBC3 and MSP 1A984; OMEGA 1287_2 and 1448_5). Offsite observation at $L_S$ 136.2° reveals $CO_2$ ice at 37.7° but not at 35°–35.5° on steep pole-facing slopes (CRISM MSP 26296). To summarize, water ice is abundant at Terra Sirenum #1 (probability level: 5) while $CO_2$ ice is not detected but with an insufficient time sampling to disprove its formation over a narrow $L_S$ range (probability level: 3).

One of the first-ever reported new bright deposits is located on the southeast facing wall of Naruko crater (36.2°S, 198.3°E) (Malin, et al., 2006). Water ice started to accumulate by $L_S$ 86.7° (CRISM MSP B21C). A 40 m/pix observation obtained when ice stability is maximum ($L_S$ 113.8°) shows $CO_2$ signatures restricted to the pole-facing wall (CRISM HRS 251E3). We detect only water ice afterward ($L_S$ 129.2° and 148°); water ice extends toward southwest and southeast facing walls (CRISM HSP 25DB0 and 26903). Ice has sublimated away by $L_S$ 160.1° (CRISM MSW 3B4F). We thus attribute a water ice level of 5 and a $CO_2$ ice level of 3 (45° difference in slope orientation but availability of high spatial sampling at appropriate $L_S$).

The next active site, "Near Gorgonum #1" (36.4°S, 190.4°E), has a southeast facing orientation. Ice is not detected offsite at $L_S$ 63.8° (CRISM FRT A9A8). At $L_S$ 115.1° we observe $CO_2$ ice on an offsite pole-facing slope and water ice on a wider slope orientation range (CRISM MSP 2887). Water ice signatures are then observed alone onsite at $L_S$ 133.7°S (CRISM HSP 260B0). $CO_2$ ice may, however, still be present as it is observed at $L_S$ 145° only 0.4° poleward (CRISM MSP D07B). Water ice may be present up to $L_S$ 156.4° (offsite CRISM HRL 390E), while no more ice is observed at $L_S$ 175.7° (CRISM MSP 1BB27).

Reported activity at Gorgonum Chaos (37.2°S, 188.3°E) is similar to that at Ariadnes Colles with numerous flows interpreted to be perennial neutral tone deposits (Dundas, et al., 2015). Flows are located on southeast and east facing slopes and form during fall or winter (Table 1). Several onsite CRISM 200 m/pixel observations are available, notably at $L_S$ 105.3° (MSP 24B97 and 2629E, HSP 26764, MSP 1AC8E and D2B7). However, steep portion of slopes in the Chaos are frequently subpixel. Both $CO_2$ and water ice are detected only once in the Chaos, at 37.6°S and $L_S$ 136.2°. Relevant offsite observations do not show ice at $L_S$ 78.4°S and 168° (OMEGA 3218_4 and 1441_5).

A new bright deposit has been observed at Pursat crater (37.4°S, 130.7°E), on the northeast facing wall, where we do not detect ice on dedicated observations obtained at $L_S$ 57.7, 101.5° and 133.1° (CRISM FRT 16E6A, HSP 19339, and 26068). The northeast wall is, however, rugged and it includes shadowed local areas that could behave similarly to the pole-facing wall of the crater. Both water and $CO_2$ ice are observed on the main pole-facing wall of the crater at $L_S$ 133.1°. Offsite data show no ice at $L_S$ 63.4°, both ices on pole-facing slopes at $L_S$ 94.2°, water ice only at $L_S$ 150.3°, and no more ice at $L_S$ 164.6° (CRISM FRT A96F and 18EDD, MSP D2BE and 3D7B). We rank this site 3 for both ices to balance the pole-facing detection of ices with the equator facing, but with local slopes, activity wall.

A new neutral-tone flow formed prior $L_S$ < 120° has been observed on a south-southwest facing crater wall of site "Terra Sirenum #2" (37.4°S and 229°E) (Dundas, et al., 2015). A fresh-looking bright deposit has also been reported on the southwest-west facing wall of this crater (McEwen, et al., 2007b). Faint spectral signatures of water ice are observed offsite at $L_S$ 61.6° and then onsite at $L_S$ 84.4° (CRISM MSP 2CB15 and B0A6). Stronger water ice signatures are observed onsite at $L_S$ 136.3° and 140.2° on large fraction of the south and southwest facing walls (CRISM HSP 300AF and 26567). Water ice is inferred to sublimate between $L_S$ 151.7° and 163° (CRISM MSP D37A and OMEGA 1404_5). We do not detect $CO_2$ ice within this crater, but early observations are not available. The next site ("Terra Sirenum #3"), nearby, however, suggests $CO_2$ ice may be present on the pole-facing wall: we have thus attributed a $CO_2$ ice probability level of 3 and 2, respectively, for the two slope orientations.

### 2.2.6 Channel incisions at 38°S

The next four sites as we move poleward have seen the formation, lengthening, or widening of channels (Dundas, et al., 2015). The season of changes is essentially unconstrained. The dominant slope orientation is pole-facing.

A major channel incision has been observed at site Terra Sirenum #3 (37.5°S, 222.9°E) (Dundas, et al., 2015). Available onsite observations do not reveal $CO_2$ ice at $L_S$ 130.8°, 138.4° or $L_S$ 143.2° (OMEGA 1177_6 at 1 km/pix, CRISM HSP 30209 and MSP 335E), while water ice signatures are abundant all over the pole-facing wall, extending toward the west facing wall. This water ice is inferred to have formed after $L_S$ 54.2° (OMEGA 567_1) and to disappear somewhere between $L_S$ 151.2° and 162.8° (CRISM MSP D336 and 3CAD). A slightly poleward observation obtained during the maximum of $CO_2$ ice stability at $L_S$ 115.5° (CRISM MSW 28AF), however, reveals clear $CO_2$ ice signatures at 38.2°S within larger water ice deposits similar to that observed onsite.

Various changes including channel incision and a deposit with visible topography have been observed in the pole-facing crater wall of site "Near Gorgonum #2" (37.6°S and 192.9°E). Onsite observations at $L_S$ 113.3° and $L_S$ 135.4° (CRISM MSP 251B1 and HSP 3000C) reveal only water ice; the noise level of these observations is, however, high (10–20%) and spatial averaging possibilities are limited over this 6 km diameter crater. In fact, we observe $CO_2$ ice signatures mixed with water ice on a larger 37.2°S pole-facing slope at $L_S$ 119.8° (CRISM MSP 19D8F) and $L_S$ 139.2° (CRISM HSP 264C3). All ice is gone by $L_S$ 177.5° (CRISM MSW 41F9) and probably by $L_S$ 163.6° (OMEGA 1408_5).

Site "Terra Sirenum #4" (38.1°, 215.9°E) corresponds to the equatorward site with reported erosion on a nonexact pole-facing slope (southwest). Water ice started to accumulate on similar slopes before $L_S$ 77.4° (CRISM HSP 17EF9). The site is observed at high resolution at $L_S$ 90.6°: at that time $CO_2$ ice is already clearly present on the south facing slope, extending toward the southwest facing slope, while water ice is no longer detected which could be due to a masking effect of the overlying $CO_2$ ice (CRISM FRT 241BB). $CO_2$ ice is still observed onsite at $L_S$ 136.2° but now with clear water ice signatures (CRISM MSP 26296). $CO_2$ ice is already gone by $L_S$ 153.3°, while water ice sublimate between $L_S$ 153.3° and 159° (offsite observations CRISM MSP D406 and 3AA7).

Site "Terra Sirenum #5" (38.1°S, 224°E) is less than 100 km away from site Terra Sirenum #3 and at the same latitude as nearby site Terra Sirenum #4. Activity is observed both on southwest (deposit) and southeast (erosion) facing slopes. Offsite observation CRISM MSW 28AF ($L_S$ 115.5°) indicates that $CO_2$ ice should be present at this time, probably also on nonexact pole-facing orientation. We observe deep water ice signatures without $CO_2$ ice onsite at $L_S$ 130.8° and 131.7° (OMEGA 1177_6 and CRISM HSP 25F58). Water ice is inferred to form between $L_S$ 49° and 79.3° and to disappear between $L_S$ 154.2° and 162.8° (CRISM FRT 16633, MSP 23757, 37CF, and 3CAD).

### 2.2.7 Penticton crater

One of the first two new reported bright gully deposits occurred on the northwest facing wall of Penticton crater (Malin, et al., 2006), at 38.4°S and 96.8°E. This site is located in the upper east part of the Hellas basins, at –2.5 km altitude, where water ice is expected to be largely stable at the surface, even on nonpole-facing slopes, according to previous studies (Langevin, et al., 2007). Indeed, water ice is largely observed within the crater, including at location of the new bright deposit on the northwest facing wall at $L_S$ 97.2° and $L_S$ 135.3° (CRISM MSP B7AE and HSP 26204). Water ice is not observed on the equator facing slope earlier and later, at $L_S$ 88.7° and $L_S$ 147° (CRISM HSP 2DEE4 and MSW 34C7). When observed ($L_S$ 97.2°), $CO_2$ ice is restricted to the pole-facing wall. Contrary to Pursat crater, we do not observe shadowed slopes due to smaller scaled topography in the vicinity of the deposit and have thus assigned a rank of 1 for $CO_2$ ice.

### 2.2.8 Simois Colles, Niquero crater and Corozal crater

Nearby sites Simois Colles (38.6°S, 183.8°E) and Niquero crater (38.7°S, 194°E) show activity restricted to deposits on southeast-facing crater walls. Limited deposits of $CO_2$ ice, without $H_2O$ ice, are already apparent at Niquero at $L_S$ 91.2° (CRISM FRT B428). Water and $CO_2$ ice are detected offsite near Simois Colles at $L_S$ 137.5°S and 141.9° and over the pole-facing wall (extending toward side walls) of Niquero crater at $L_S$ 151° (CRISM HSP 1A7B8 and OMEGA 1258_3, CRISM MSW 3658). No more ice is seen at these latitudes at $L_S$ 163.6° (OMEGA 1408_5).

A new dark deposit and a slight channel incision occurred, possibly between $L_S$ 61° and 183°, within a gully of the south/southwest facing wall of Corozal crater, at 38.8°S and 159.5°E (Dundas, et al., 2010). $CO_2$ ice is detected onsite at $L_S$ 106.6°, with faint signatures of water ice (CRISM MSP 195FA). Offsite, $CO_2$ and water ice are then observed at $L_S$ 134.9°, 140.6°, and 153.2° but not anymore at $L_S$ 172.3° (CRISM MSP 2FFBF, 1A991, 3768, and 1B98C).

### 2.2.9 Unnamed crater in Terra Sirenum

A new small-scaled channel incision has been observed at site "Terra Sirenum #6" (38.9°S, 223.7°E) (Dundas, et al., 2012). The authors indicate that this change occurred between $L_S$ 136° and 158°, with an illustrative figure showing $L_S$ 44° and 158° observations only. As discussed in a preliminary analysis (Vincendon, et al., 2013), this difference is of importance when looking at ice composition. We have performed our own investigation of the seven available HiRISE images of site obtained at $L_S$ 44°, 95°, 115°, 122°, 128°, 136°, and 158°: a dark deposit indeed appeared between $L_S$ 136° and 158° (C. Dundas, personal communication); however, we are not able to confirm the $L_S$ 136°–158° range for the channel incision due to the lower spatial resolution of HiRISE images obtained between $L_S$ 95° and 136°.

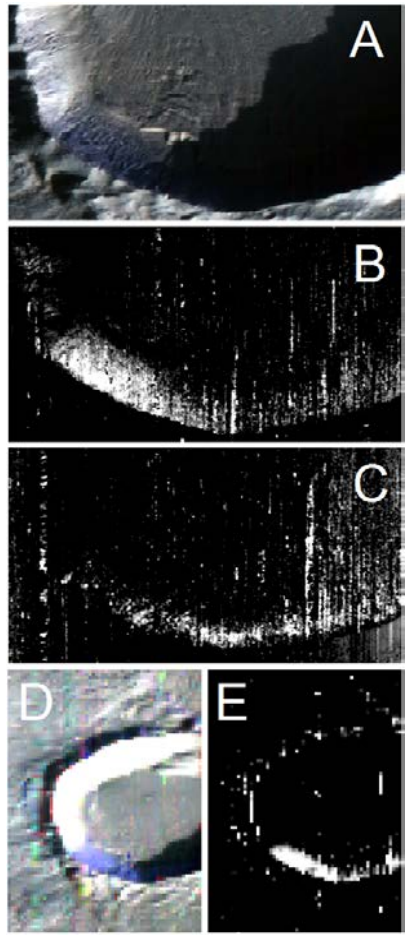

Figure 4: Examples of near-IR observations obtained for site Terra Sirenum #6 (38.9°S, 223.7°E). Top (A, B, C): CRISM FRT 248D4 ($L_S$ 101.1°, 20 m/pix). Bottom (D, E): CRISM HSP 1A709 ($L_S$ 136°, 200 m/pix). Figures 4A and 4C are near-IR reflectance RGB views (associated wavelengths: 2.5 µm, 1.5 µm and 1.1 µm respectively). Raw water ice 1.5 µm band depth maps with intensities from (Figure 4B) 2 to 12% and (Figure 4E) 10 to 20%. Figure 4C is the $CO_2$ 1.43 µm raw band depth map, from 6% to 15%. See (Pelkey, et al., 2007) for details about CRISM automatic spectral index mapping.

Water ice started to accumulate between $L_S$ 44° and $L_S$ 74.4° and $CO_2$ ice between $L_S$ 74.4° and 93.8° (CRISM HSP 161B8, 23395 and 243EB). Onsite data reveal that $CO_2$ ice sublimated somewhere between $L_S$ 122.6° and 130.8°, while water ice sublimated between $L_S$ 151.2° and 174.1° (CRISM MSP 19EBC, OMEGA 1177_6, HSP 1A709, MSP D336 and HSP 26A52). Offsite data make it possible to narrow this range to $L_S$ 154.2° - 162.8° (CRISM MSP 37CF and 3CAD). Ices are observed exactly at the location of the gully activity (Figure 4), on the southwest facing wall. These observations, as well as others not detailed here, are used to construct a precise timing of ice formation and disappearance at this site (Figure 5). The $L_S$ 136°–158° time range, which corresponds to the dark deposit formation and possibly to the channel incision, includes only water ice; the more conservative $L_S$ 54°–158° time range for channel incision includes both ices. $CO_2$ ice probability level is thus 1 for the dark deposit and 3 for the channel incision, while water ice probability levels are 5 for both ices.

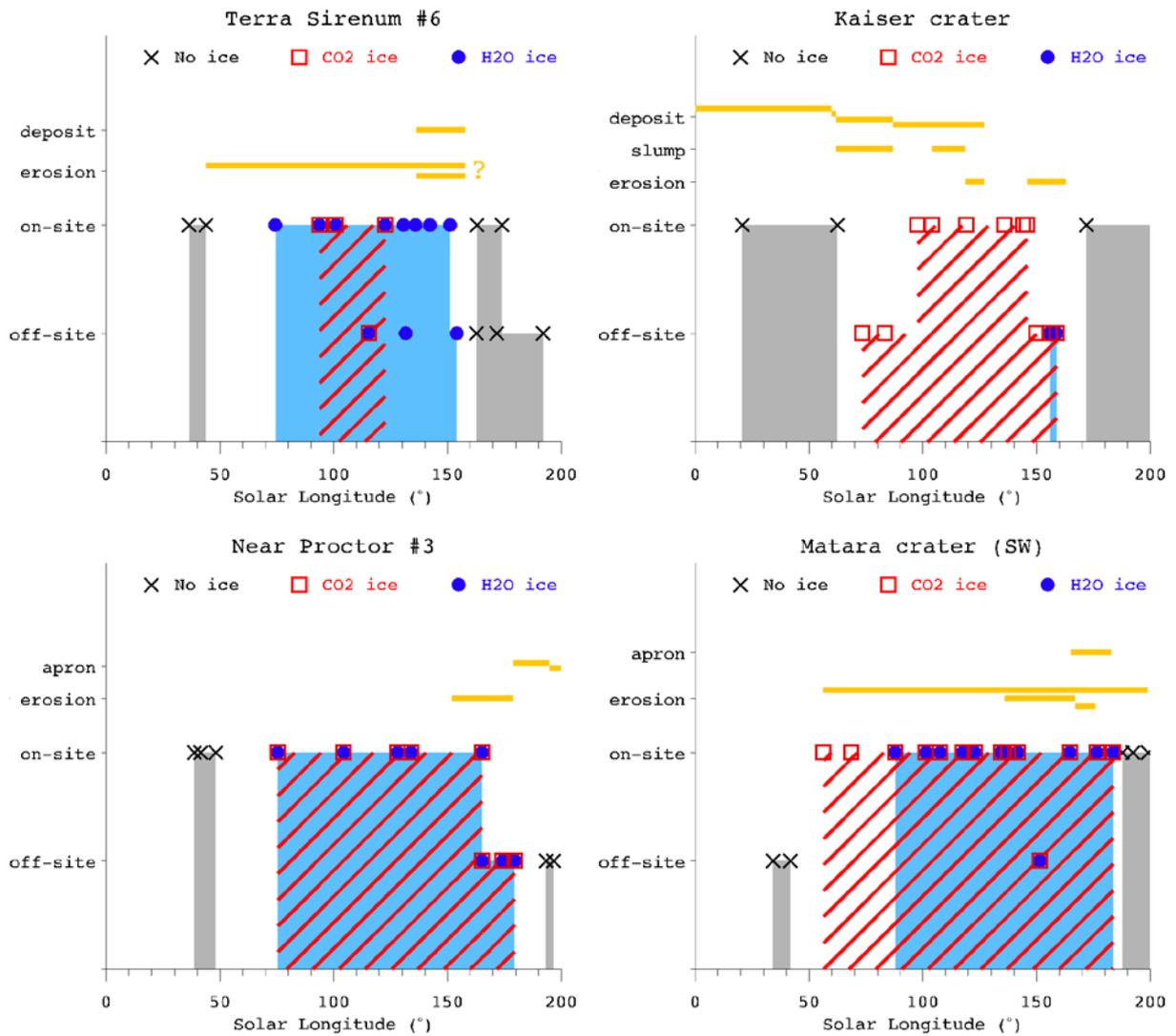

**Figure 5:** Same as Figure 3 for sites Terra Sirenum #6, Kaiser crater, Near Proctor #3, and Matara crater (see Table I for details).

### 2.2.10 Activity about 40°S

The last active site reported equatorward 40°S is within Arrhenius crater at 39.6°S and 123.1°E, with a new deposit observed between $L_S$ 100° and $L_S$ 180°. Near-IR data are restricted to offsite observations. We infer that water ice formed at $L_S$ < 83.7° (CRISM HSP 15C7F) and sublimate away between $L_S$ 152.2° and 175.9° (CRISM MSP 36F7 and 4152). $CO_2$ ice may be already present at $L_S$ 83.7°; it is clearly observed at $L_S$ 133.9° (OMEGA 1200_5) and it probably sublimates between $L_S$ 145.3° and 149.5° (CRISM MSP 1AC2F and 35D5). Both ices are observed at southwest facing orientation, e.g. at $L_S$ 133.9°S.

A "topographic change in channel" (Dundas, et al., 2015) has been observed at 40.3°S and 217.1°E (site "Terra Sirenum #7") on the gentle slopes (about 20° maximum) of a southwest facing crater wall. Onsite, we do not yet detect ice at $L_S$ 76.2°S while we observe a faint signature of water ice at $L_S$ 82.2° (CRISM MSP 234CB and HSP 2DA27). Strong signatures of $CO_2$ ice without clear evidence for water ice are then observed at $L_S$ 89.5°, extending toward southwest facing orientation (CRISM FRT 18BBC). Water ice, with $CO_2$ ice, is observed again offsite at $L_S$ 133.7° (CRISM MSP 260AB). Onsite observations then reveal water ice only at $L_S$ 141.4° and 153.3° and then no more ice at $L_S$ 165° (OMEGA 1254_3 at 1.5 km/pix, CRISM MSP D406, and 3DB8).

The next site is located within Argyre planitia at 40.5°S and 309.9°E on a sandy slope. The indicated location (40.5°S and 309.9°E) suggests a south-southeast facing orientation. An offsite observation reveals clear signatures of both $CO_2$ and water ice at $L_S$ 103.8° (CRISM HSP 24AA6).

### 2.2.11 Avire crater

Numerous new linear channels were observed on the southwest facing sandy wall of Avire crater (40.8°S, 200.3°E). $CO_2$ ice without detectable contamination by water ice is observed on the southwest facing wall at $L_S$ 96.6° (CRISM FRT 2458F). Subsequent observations show both water and $CO_2$ ice signatures at $L_S$ 134.7° and 142.8° (CRISM HSP 26193 and 266C9). Offsite observations suggest that both ices sublimate between $L_S$ 147.3° and 157.4° (CRISM HSP 1AD1B and MSP 39BF).

The two nearby sites "Near Avire #1" (41.1°S, 203.5°E) and "Near Avire #2" (41.1°S and 189°E) are each covered by a single onsite observation obtained at $L_S$ ~ 146° showing water and $CO_2$ ice (CRISM HSP 2680F and MSP 1AC8E). Changes occurred on the southwest facing crater walls. Note that for the three Avire sites, ice is not detected on the southeast facing wall while it is on the southwest-facing one.

### 2.2.12 Palikir and Roseau craters

Channel widening and deposition of material occurred in a gully of the southwest facing wall of Palikir (41.4°S, 202.3°E) during fall or winter (Dundas, et al., 2012). Ice is not yet detectable at $L_S$ 84.4° (offsite CRISM MSP B0AB). Strong signatures of $CO_2$ ice without significant evidence for water ice are then observed all over the southwest to southeast facing wall of the crater at $L_S$ 113.7° (CRISM MSP 27EF). We still observe $CO_2$ ice offsite at $L_S$ 136.5°, with faint signatures of water ice (CRISM MSP 1A74E). $CO_2$ signatures are stronger and are detectable longer on the southwest facing wall compared to the southeast one. No more ice is observed onsite at $L_S$ 152.3° (OMEGA 1331_4, 2 km/pixel). $CO_2$ ice may already have sublimated by $L_S$ 148.5°, while water ice is probably still present at $L_S$ 145.9° (offsite CRISM HSP 26957 and 2680F).

A bright deposit formed between $L_S$ 354° and 167° on the east facing wall of Roseau crater (41.7°S, 150.6°E). We do not detect ice within the crater at $L_S$ 71.6° (CRISM HSP 179DD) and then at $L_S$ 81.6° and 84.3° with 1.6 km/pix OMEGA observations (3244_5 and 3266_5). Contrary to Palikir, both water and $CO_2$ ice are largely observed on the south and southwest facing wall of the crater at $L_S$ 137.8° and 145.2° (CRISM HSP 263CF and 1AC28). However, there are no spectral detections of ice on the east, or even southeast, facing wall of the crater, where activity does occur, which lead us to assign a probability level of 3 for both ices.

### 2.2.13 Kaiser crater

Both dark deposits and channel incisions were observed on the west facing slope of a dune in Kaiser crater (46.7°S, 20.1°E) (Diniega, et al., 2010; Dundas, et al., 2012). The dark flows appear regularly between $L_S$ 324° and 127° and could be transient flow over ice and/or somehow more perennial ground deposits (Pasquon, et al., 2015). Major channel formation is observed between $L_S$ 119° and 127° followed by minor erosion between $L_S$ 146° and 163°. Visible imagery suggests frost between $L_S$ 60° and 172° (Dundas, et al., 2012). We do not observe spectral evidence for ice on the high spatial resolution onsite observations obtained at $L_S$ 20.6° and 62.4° (CRISM FRT 98ED and 17220). $CO_2$ ice is not expected at these solar longitudes for this latitude (Vincendon, et al., 2010a), even more on a west facing slope, and expected water ice thickness are limited to 5 µm maximum at that time (Vincendon et al. 2010b). We detect $CO_2$ ice with no spectral evidence for water ice offsite at $L_S$ 73.4° (OMEGA # 8016_0, > 10 km/pix) and $L_S$ 83.5° (OMEGA # 3260_5, 1.5 km/pix). This last OMEGA observation shows that water ice is observed between 30°S and 40°S but not poleward; this atypical behavior is expected at this longitude due to atmospheric flow of dry air (Vincendon, et al., 2010b). Two noisy 200 m/pix observations have been obtained onsite at winter solstice ($L_S$ 86.6° and 87.7°, CRISM HSP 18910 and 23F7F); the west facing side of the dune is only 3–4 pixels wide and the resulting noise after averaging is still 10–20%: ice is not identified in these conditions. Then, $CO_2$ ice, without significant spectral signatures of water ice, is regularly observed onsite at $L_S$ 97.7°, 104°, 119.1°, 135.9°, 144° and 146° (CRISM FRT 2461A, MSP 194BC and 19D30, HSP 26270 and 26746, and MSP 1AC74). Offsite, $CO_2$ ice signatures are still observed at $L_S$ 149.9° (CRISM MSP D29A), while water ice is then observed in addition to $CO_2$ ice at $L_S$ 155.8° and 158.8° (CRISM D524 and 3A85). No more ice is detected onsite at $L_S$ 171.9° (MSP 1B970). These constraints are compared to reported change activity in Figure 5. Formation of new channels/channel lengthening and widening ("erosion"), and observation of slumping material movements ("slump") all occur in time range where $CO_2$ ice is observed. Dark lineae ("deposit") occurring at $L_S$ lower or equal to 60° are not concurrent with ice deposits detectable by CRISM or OMEGA.

### 2.2.14 Asimov crater

Bright, fresh looking deposits have been observed on a northwest facing crater wall at 46.9°S and 4.3°E within Asimov crater (McEwen, et al., 2007b). Activity with new dark deposits has then been observed at this location (Dundas, et al., 2015). The timing of activity is unknown. We do not observe $CO_2$ nor water ice onsite on the northwest facing wall at $L_S$ 146.1° and 165.9° (CRISM MSW 3470 and FRT 3E23). Strong signatures of water and $CO_2$ ice are observed but on other slope orientations: pole-facing at $L_S$ 165.9°, largely extending to southwest facing at $L_S$ 146.1°. Offsite observations obtained at $L_S$ 103.7° and 118.5° (CRISM HSP 24A99 and OMEGA 1084_3 at 1.2 km/pix) do not reveal ice on northwest facing slopes; $CO_2$ ice without significant water ice signatures is however observed to extend on flat surfaces at that times. There is no ice on any slope orientation at $L_S$ 37.1° (CRISM MSW A1F2) and $L_S$ 187.3° (OMEGA 1560_2). There is thus no indication for onsite ice, but nearby signatures of ice are abundant.

### 2.2.15 Proctor crater

Activity has been reported on dune fields and sandy slopes located in Proctor and surrounding smaller craters, west of Hellas basin (Diniega, et al., 2010; Dundas, et al., 2012; Dundas, et al., 2015).

"Near Proctor site #1" is located at 36.7°E, 45.8°S. No information is available, although coordinates suggest a pole-facing orientation. Both $CO_2$ and $H_2O$ ice are observed on the pole-facing slope at $L_S$ 147.5° (CRISM onsite HSP 1AD36). The two ices disappeared between $L_S$ 166.4° and 177.9° (CRISM observations offsite MSP 3E56 and onsite HSP 26C13) and are not yet detected at $L_S$ 66.4° (onsite observation CRISM HSP 1755C).

Multiple sites with activity are reported for the Proctor crater dune field centered about 47.5°S, 30.4°E (Dundas, et al., 2015). The formation of a new channel was, e.g., observed on a southwest/west facing wall at 47.8°S, 30.7°E (Diniega, et al., 2010). We detect both ices all over the dune field during winter at $L_S$ 104.9°, 128.1°, and 141.3° (CRISM MSP 1953F, OMEGA 1157_7 1.8 km/pix, and CRISM MSP 1AA03). The spatial resolution of available observations is insufficient to characterize precisely the spatial distribution of ice over a given dune. No ice is observed over the dune field at $L_S$ 48.1° (OMEGA 5418_7, 4 km/pix) and $L_S$ 177.7° (CRISM MSW 4221). Both ices are probably still present onsite on southwest facing slopes at $L_S$ 162.6° (CRISM MSP 3C83).

Dune activity of unknown nature and timing is also reported in another crater just east of Proctor crater at 47.2°S and 34°E (site "Near Proctor #2"). Ice formation within this dune field is compatible with that of Proctor crater, with some additional constraints available to assess the beginning and ending of the icy period. Both water and $CO_2$ ice are observed onsite at $L_S$ 74.6°, 102.6°, 144.4°, and 160.1° (CRISM MSP 233BE, HSP 19405, MSP 1ABC1, D72C). Onsite observations at $L_S$ 172° and 174.4° do not reveal ice spectral features within the dune field (CRISM MSP 3FC2 and 26A76). Both ices are, however, observed slightly equatorward on large pole-facing slopes. Offsite observations suggest ice has not yet started to significantly accumulate at $L_S$ 56.4° (CRISM MSP 2C7E2) and has disappeared by $L_S$ 186.6° (OMEGA 1556_2).

"Near Proctor site #3" (49°S, 27.2°E) is located poleward: activity has been reported on a pole-facing sandy 20 km diameter crater wall over $L_S$ 152°–195° essentially. Ice is not yet observed onsite and around at $L_S$ 48.1° (OMEGA 5418_7, 3.7 km/pix). Both ices are then observed onsite at $L_S$ 75.4°, 104.4°, 128.1°, 134.1°, and 165.3° (OMEGA 3194_6, CRISM MSP 194F6, OMEGA 1157_7 and 1201_4, CRISM MSP 3DDC). Offsite observations then suggest that both ices sublimate somewhere between $L_S$ 179.6° and $L_S$ 193.3° (CRISM HSP 26CAC and MSP 10930). These observations and additional ones are compared to reported change activity in Figure 5.

Finally, another active dune site is reported at similar latitudes (48°S) but west of Argyre basin (303.7°E) on the west-southwest facing side of the dune (C. Dundas, personal communication), without available timing constraints. The west-southwest facing slope is largely covered by $CO_2$ and water ice during winter (onsite CRISM HSP 24201 and 26789 at $L_S$ 91° and 144.7°). This ice is probably already present at $L_S$ 66.7° (CRISM high-resolution but visible-only onsite observation FRT 22E70).

### 2.2.16 Matara crater

The dune field within Matara crater (49.5°S and 34.8°, near Proctor crater) shows extensive gully activity (Diniega, et al., 2010; Dundas, et al., 2012). Gullies formation and modification are particularly impressive with change observed every year at several locations of the dune field. Major changes (new channels and alcove expansion) are observed on southwest-facing slopes. Small-scaled erosion is also observed on northeast facing slopes within alcoves (Dundas, et al., 2015). We have extracted in Table I the major typical activity periods; however, additional active events are also indicated in the publications referred previously.

About 20 onsite observations are available over fall and winter: we mention only the key ones here while most of these observations are used to construct the time diagram of Figure 5. Ice appears between $L_S$ 41.9° and 56.5° (CRISM HSP 15F23 and FRT 16D30). Up to $L_S$ 68.6° (CRISM HSP 17742) we detect only $CO_2$ ice on pole-facing slopes extending toward southwest and southeast orientations. At $L_S$ 88° (CRISM HSP 18A78) ice extends to flat surfaces and water ice spectral signatures are detected. Between $L_S$ 101.4° and 134.4° (CRISM MSP 248FC and HSP 26160) both ices are observed with increasing band depth intensity: ice extends to the bottom part of the equator facing crater wall with a gentle slope angle (< 5–10°), while being not stable on the major, steeper portion of the wall. The spatial extent of ice start to decrease about $L_S$ 137.4°/141.8° (CRISM HSP 26398/MSP 1AA59) and ice is no longer detected on flat surfaces at $L_S$ 151.5° (CRISM MSP D35D). Ice remains clearly observed on southwest facing slopes at $L_S$ 164.7° and 176.5° (CRISM MSP 1B6A6 and HSP 1BBBB). Localized ice patches are still detected at $L_S$ 183.7° (CRISM HSP 26E71), while ice is not observed anymore at $L_S$ 187.7° (CRISM MSP 460D).

# 3 Discussion

## 3.1 Observational biases

The main limitation of our near-IR dataset is linked with the spatial sampling of the available CRISM and OMEGA observations, ranging from 20 m/pix to several km/pix: changes can be a few meters wide only (e.g., Terra Sirenum #6), and are in any case frequently either subpixels (e.g., in Ariadnes Colles) or covered by a few pixels at best. The upslope initiation area of some change might be even smaller. In such cases ice will be detected in the near-IR observations if it covers globally the area where change is observed, or if subpixel deposits are sufficiently large, or sufficiently thick and/or large grained, so that the resulting band depth after averaging with non-icy material of the pixel can be detected. In certain circumstances, it is possible to enhance detection capabilities by carefully selecting and averaging relevant near-IR data pixels based on slope properties and data quality considerations, as illustrated at Ariadnes Colles (see section 2.2.3). Looking for ice offsite, where slopes of a given orientation can be significantly larger, also proved to be a useful way to overcome some of the limitations linked with spatial resolution (e.g., Near Gorgonum #2). In other cases, the lack of high-resolution data and relevant offsite observations is suspected to prevent ice detection (e.g. at Kandi crater). Overall, we thus evaluate in this study a presence of ice compatible with the change, which can be exactly on the change itself or in the immediate vicinity (e.g., on the steep slope overhanging a new deposit).

Ice detection is also clearly prevented in some case by the low signal-to-noise ratio of available data, notably for small-scale sites with limited pixel averaging possibilities. In good observational conditions, water ice band depths greater than 4%, corresponding to water ice thickness 2 to 5 µm thick, should be detectable (Vincendon, et al., 2010b). Longer photon path lengths are required within $CO_2$ ice (typically of few hundreds of µm) due to the presence of $CO_2$ gas in the atmosphere (Langevin, et al., 2007; Appéré, et al., 2011). However, $CO_2$ is the main component of the atmosphere and $CO_2$ ice layers rapidly reach this thickness once condensation starts. Observations of interest mostly concern pole-facing slopes near winter solstice, when the reflectance measured from the surface is insufficient to reach nominal CRISM or OMEGA performances. Random noise can reach 20% of signal level while ice features are observed to be limited to 5% at some sites (see Figure 1). At Kaiser crater (see section 2.2.13), we observe that $CO_2$ ice can be undetectable in winter solstice observations while being observed before and after, and we do not detect water ice in autumn while very thin, about 5 µm thick deposits are expected and probably seen with high-resolution visible imagery (Dundas, et al., 2012).

The maximum ice stability is reached between $L_S$ 100° and $L_S$ 130° at low latitudes (Vincendon, et al., 2010a; Vincendon, et al., 2010b) and later on ($L_S$ 130–150°) at higher latitudes (Kereszturi, et al., 2011). Missing observations over these timeframe can thus prevent ice detection. Ice needs to accumulate several days to be detectable at near-IR wavelengths, so the local time of observation during the day does not influence results (Vincendon, et al., 2010a; Vincendon, et al., 2010b). We must, however, note that very thin (μm, maybe tens of μm thick for $CO_2$ ice) transient layers of frost forming at night could be present and will remain undetected.

The superimposition of ice layers with variable compositions can alter ice identification. At some sites such as Terra Sirenum #4 and #7 we observe first the formation of a water ice layer which is then covered by a $CO_2$ ice layer sufficiently thick, small-grained, and/or minimally contaminated by water ice so that water ice features are no longer detectable. Water ice signatures are observed again afterward, as water ice accumulates while $CO_2$ ice stops growing or sublimates. The reverse spectral masking effect of water ice above $CO_2$ ice is not suspected in our study, notably because we never observe the reappearance of $CO_2$ ice spectral features. This effect has been observed and modeled at polar latitudes (Appéré, et al., 2011). Modeling suggests that a 50–100 μm thick water ice layer, which typically corresponds to the total amount of water ice forming at our latitudes (Vincendon, et al., 2010b), could be enough to hide an underlying $CO_2$ ice layer (Appéré, et al., 2011). The accumulation of water ice is, however, progressive and we do not expect that such a quantity could form after $CO_2$ ice stops growing. Likewise, if $CO_2$ ice sublimates while water ice remains, we expect from these rough estimates that approximately all $CO_2$ ice need to sublimate to get a sufficient masking thickness of water ice.

We have limited our study to sites for which current activity has been securely reported (Dundas, et al., 2015). At two sites, we also discussed and included in diagrams "fresh looking" bright deposits observed prior activity (McEwen, et al., 2007b). These two sites (Asimov crater and Terra Sirenum #2) were targeted due to these bright deposits, but observed activity is slightly different (different color and/or orientation). We included these two bright deposits because they are similar to the first two new deposits reported (Malin, et al., 2006) as well as to that observed at Roseau crater (McEwen, et al., 2007b) where subsequent observed activity was also a bright deposit (Dundas, et al., 2010). Another potential bias in our understanding of the activity data set concern some dark deposits, as it is not always clear whether they are perennial or ephemeral surface deposits, or transient phenomena occurring above ice, but dissipating afterward (Pasquon, et al., 2015). This notably concerns sites Ariadnes Colles, Gorgonum Chaos, and Kaiser crater.

## 3.2 Overview of ice distribution and properties

We have conducted a study of ice formation site by site. In agreement with previous global studies (Vincendon, et al., 2010a; Vincendon, et al., 2010b), we observe that the pattern of ice formation depends to first order on latitude but with some significant regional longitudinal variations.

While the equatorward deposits of $CO_2$ ice reported so far were located at about 34.5°S (Vincendon, et al., 2010a), we have detected $CO_2$ ice twice ($L_s$ 90° and 100°) at a latitude of 32.3° on a restricted portion of the steep pole-facing wall of site Promethei Terra, east of Hellas (118.6°E). The frost point of $CO_2$ ice increases with pressure. Indeed, $CO_2$ ice was previously not observed equatorward 45°S in the high-altitude area of Solis Planum (Vincendon, et al., 2010a). We can notice that gullies are less abundant in that area (Harrison, et al., 2015), with no reported activity (Dundas, et al., 2015). The four equatorward detections of $CO_2$ ice reported so far all occur north of Argyre or east of Hellas which correspond to the two regionally low-altitude areas of this latitude band (the Hellas basin itself placed apart). $CO_2$ ice detections are restricted to exact pole-facing orientation for latitudes equatorward 38°S.

While $CO_2$, the main component of Mars atmosphere, is always available to condense once surface temperatures reach frost point, water is a minor component of the atmosphere and its formation at the surface is a more complex phenomena which depends on surface temperature, water vapor pressure, and snowfall (Vincendon, et al., 2010b). Overall, the accumulation of surface water ice generally occurs at warmer temperature compared to $CO_2$ ice. This gives an advantage to the formation of water ice for active sites located equatorward of 40°S: water ice generally formed earlier, sublimated later, and formed on a wider range of slope angle orientation. As we approach the seasonal cap at poleward latitudes (≥ 45°S), we observe more identical timing and spatial distribution for both ices, although longitudinal variations are observed (Figure 5) with, e.g., a lower presence of water ice in the dry area of western Noachis: we do not detect water ice at site Noachis Terra #1 and water ice accumulation remains long insufficient to be detectable at Kaiser crater.

The timing of ice formation and disappearance at the surface provides elements to constrain the thickness of the ice layers that accumulate on the ground. We used the 1D energy balance code developed at the *Laboratoire de Météorologie Dynamique* in Paris (Forget, et al., 1999; Millour, et al., 2009) and modified to simulate ice condensation on slopes (Spiga & Forget, 2008; Forget, et al., 2008; Vincendon, et al., 2010a; Vincendon, et al., 2010b) to estimate ice thickness. Ice timing constrained by near-IR spectroscopy is compared to model predicted stability timeframes. As discussed by (Vincendon, et al., 2010a), predicted $CO_2$ ice stability depends on model parameters such as ground thermal inertia, ice albedo, subsurface ice depth, etc. These parameters are not strongly constrained by independent measurements, especially on crater walls: they can be tuned to explore possible $CO_2$ ice formation timing. We can then estimate a range of expected $CO_2$ ice thicknesses by comparing model outputs to the observed timing and their uncertainties. Model parameter dependence is lower for water ice, and we derive consistent result with the model as expected from earlier work (Vincendon, et al., 2010b). These evaluations only provide an order of magnitude of the thickness: as potential uncertainty sources and biases are numerous, we do not aim at providing a comprehensive evaluation of the uncertainties linked with these retrievals. With this method we infer a $CO_2$ ice layer between 2 and 17 mm thick at Promethei Terra, the most equatorward site for which $CO_2$ ice has been reported (Figure 6). A thickness lower than 10 mm is inferred for site Terra Sirenum #3, the equatorward site (37.5°S) where an imprecise new channel formation has been observed (Dundas, et al., 2015). Similarly, at Terra Sirenum #6 (38.9°S), where a small channel incision has been observed, modeled $CO_2$ ice thicknesses are a few millimeter thick at maximum (Vincendon, et al., 2014). At Gasa crater, where timing constraints are precise, we model a $CO_2$ ice thickness of 9 ± 2 mm. For a given longitude, the thickness of $CO_2$ ice generally increases as we move poleward. However, there are some longitudinal variations linked notably with pressure (Vincendon, et al., 2010a) which can result in significant site-to-site differences for latitudes close to ice stability. Modeled $CO_2$ ice thicknesses become clearly larger poleward of 45°S where long $CO_2$ ice timeframes are observed: the thickness is e.g. inferred to be 4 ± 1 cm on Kaiser crater (46.7°S) steep pole-facing slopes.

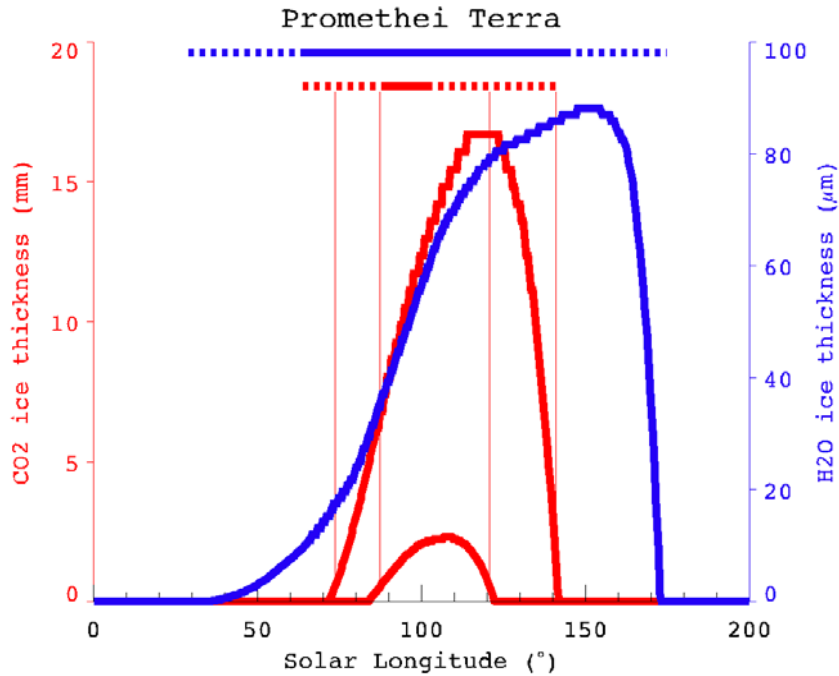

**Figure 6: Modeled ice thicknesses that accumulate on the 30° pole-facing slope of Promethei Terra (32.3°S, 118.6°S). Near-IR observational constraints about the presence of ice are summarized on top (horizontal lines). Two scenarios that frame the observed uncertainty range (dotted lines) for $CO_2$ ice (red) are indicated: they provide maximum (17 mm) and minimum (2 mm) possible $CO_2$ ice thicknesses. Both scenarios lead to a unique similar water ice (blue) accumulation history that is compatible with observational constraints. A detectability limit of 500 µm thick for $CO_2$ ice is assumed (Appéré, et al., 2011). The detectability limit of water ice is expected to be 5 µm thick (Vincendon, et al., 2010b).**

## 3.3 Correlations between ice and activity

The empirical ice probability level estimates (see section 2.1) are referenced in Table I site by site (and activity by activity when there are enough available details). Probability levels account for the observational biases identified and discussed in section 3.1. We estimate that the uncertainty on these subjective probability levels that range from 1 to 5 is typically either +1 or –1. However, as discussed previously, the extent of potential biases frequently prevented us from reaching definitive conclusions for a single site. The analysis is thus now conducted at global scale with a statistical study of the whole sample: we compare in Figure 7 water and $CO_2$ ice probability levels at active gully sites, as a function of activity type. Obvious correlations between activity types and ice probabilities emerge from these histograms.

The main erosive modifications (formation of new channels, channels lengthening or widening, formation of new alcove and major alcove enlargement) all occur where (and, if applicable, when) $CO_2$ ice probabilities are maximum (13 sites with level 5, two with level 4 and one with level 3). This is not the case for less impressive gully changes (minor alcove or apron modifications and formation of new deposits only): nine of these 35 activities have $CO_2$ ice probability levels of only 1 or 2, with notably five sites with new bright or dark deposits that occur where and/or when no $CO_2$ ice is observed nor expected (Kibuye, Gasa, Penticton, Terra Sirenum #6, and Kaiser).

The detectable presence of water ice is compatible with most changes, with the exception of Gasa crater minor alcove modification and most Kaiser Crater activity where water ice is probably present but in low quantity (≤ 5 µm thick layer). Two other sites (color deposit and apron change) have a water ice probability level of only 2. A specific correlation emerges for bright deposits: bright deposits are anticorrelated with $CO_2$ ice while being correlated with water ice (Figure 7). This correlation must be handled carefully considering the small sample size (eight sites, two of which have a specific status, see section 3.1). This behavior is mainly due to slope orientation, as most of these bright deposits are observed on side or even equator facing crater walls. It is linked with timing constraints for Gasa crater: various deposits are observed at Gasa, but the bright deposit is the only one that is not compatible with $CO_2$ ice.

Finally, only three activities took place where or when evidence for both ices is weak (probability levels of 1 or 2): the alcove change of Gasa for which ice can be definitely excluded ($L_S > 212°$), the early dark lineae of Kaiser crater ($L_S < 60°$) and the late ($L_S > 195°$) apron changes of site Near Proctor #3. It should be noted that very thin, undetectable water ice layers (< 5 µm) could occur at Kaiser crater between $L_S \sim 25°$ and $L_S$ 65° (Vincendon, et al., 2010b).

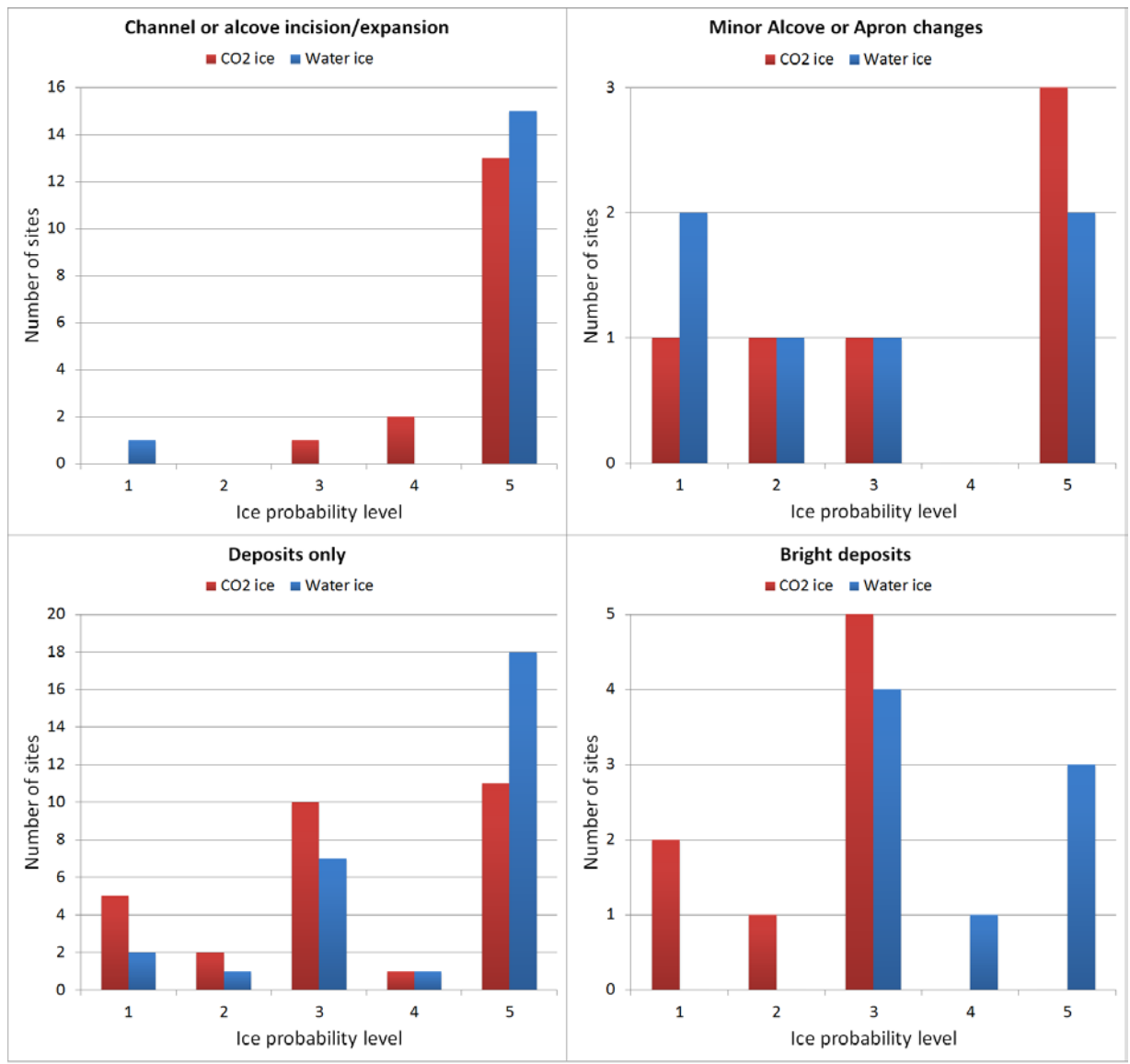

**Figure 7 : Histograms providing the number of sites/activities as a function of ice probability levels (red for $CO_2$ and blue for water). Probability level ranges from 1 (low) to 5 (high); see section 2.1 for details. Points in the diagram are linked with lines of Table I: a given site can thus contribute by several points. Four histograms are shown for four different classes of changes (see details in Table I). The "deposits only" histogram includes the "bright deposits" one; it corresponds to lines of Table I with empty erosion column cells. The five lines of Table I without indications about the change type are not included in these diagrams.**

## 3.4 Implication for gully formation

The strong correlation between $CO_2$ ice and major gully activity, including all examples of new channel formation, strongly support the idea that current gully channel formation is mainly driven by $CO_2$ ice, as suggested over the last years from HiRISE analyses (Dundas, et al., 2010; Diniega, et al., 2010; Dundas, et al., 2012; Dundas, et al., 2015). The time range of $CO_2$ ice stability compared to climate modeling prediction (section 3.2) indicates that the amount of $CO_2$ ice present at low-latitude sites is probably low with $CO_2$ ice thicknesses ranging typically from 1 mm to 1 cm. These values are compatible with observed $CO_2$ ice band depths (Figure 1, Figure 2) compared to synthetic spectra (Appéré, et al., 2011), even though assessing the link between both properties is a thorny issue. At low latitudes, $CO_2$ ice is frequently observed to be restricted to the upper, steeper part of pole-facing slopes. $CO_2$ ice is even not observed directly at Terra Sirenum #3 (37.5°S) where a major channel incision has been recently reported (Dundas, et al., 2015). These observational constraints provide inputs for deciphering $CO_2$-based formation mechanisms. Some studied mechanisms currently require the presence of large $CO_2$ blocks (Diniega, et al., 2013): these scenarios may not be fully compatible with erosion activity occurring equatorward 40°S. When change timing is sufficiently constrained, we observe that erosive changes preferentially occur late in winter season, when $CO_2$ ice is not accumulating anymore but on the contrary sublimating ("Gasa crater," "Kaiser crater," "Near Proctor #3" and "Matara crater"), which is consistent with some proposed mechanisms that do not particularly require large thicknesses of $CO_2$ ice (Hugenholtz, 2008; Cedillo-Flores, et al., 2011; Pilorget & Forget, 2015).

Water ice is almost always observed with $CO_2$ ice at active sites with erosion. At high latitude, water ice is probably mostly present as inclusions within the $CO_2$ ice matrix. At low latitude, water accumulates prior $CO_2$ ice and remains up to several tens of solar longitude degrees afterward. While we do not observe directly $CO_2$ ice at Terra Sirenum #3, we largely detect water ice. We must also notice that the channel incision of site Terra Sirenum #6 could have occurred while water ice only was onsite, if the narrow time range suggested by (Dundas, et al., 2012) is confirmed. Thus, while our findings argue in favor of a dominant role of $CO_2$ ice in current channel incision, it does not rule out a possible contribution of water ice-related mechanisms. In fact, some proposed mechanisms only require the presence of frost, whatever its composition (Hugenholtz, 2008). $CO_2$ ice could be favored over water ice simply because it is more abundant.

Activity limited to the formation of new deposits of various tones or minor topographical change (thick deposit, slumping material, and minor alcove erosion of apron change) occur preferentially in association with ice, but not necessarily $CO_2$ ice. Activity uncorrelated to the presence of ice presence seems possible but marginal. It is probably restricted to some minor collapse or slump occurring in spring after more important winter changes (Gasa, Kaiser) or to dark deposits that may result from processes similar to that causing dark streaks or recurring slope lineae (Kaiser crater). A significant contribution of $CO_2$ ice is not supported by our observations for several new deposits, such as in Kibuye crater at 29°S. We must however notice that transient thin $CO_2$ ice layers forming at night cannot be excluded with our observational constraints. Palikir and Roseau craters, located at the same latitude, provide an interesting overview of the correlation between ice and activity type. A channel widening is observed at Palikir crater where $CO_2$ ice is observed while observations indicate a weak presence of water ice. At Roseau crater, water ice form largely and the observed change is a new bright deposit. In fact, a strong anticorrelation with $CO_2$ ice is observed for all new bright deposits, while these bright deposits correlate with the presence of water ice (Figure 7). Previous studies about these bright deposits have not yet concluded in favor of dry or wet formation mechanisms (Pelletier, et al., 2008; Heldmann, et al., 2010). It has been suggested that deposit tones may just reflect ambient material (McEwen, et al., 2007b) or grain size and may not carry information about processes (Dundas, et al., 2015). A brighter material could, however, imply a process reducing grain size via, e.g., a removal of material consolidation or a process that is only effective on fine-grained material (C. Dundas, personal communication), while darkening on Mars usually evokes a process removing a fine coating of bright dust such as for dark streaks. The only deposit for which a time constraint is available occurred at Gasa crater. The bright deposit appeared over a short late winter time range that includes the presence and then sublimation of water ice. We can also notice that bright deposits are observed on nonpole-facing slopes for other sites, where surface temperatures are warmer. The formation of liquid water from ice has been hypothesized at small-scale from numerical modeling (Mohlmann, 2010). Processes involving the sublimation of water ice, with a transient unstable liquid stage, have been recently shown to create downslope movements in Martian conditions (Massé, et al., 2014). Resulting flows would not necessarily be wet: such mechanism could destabilize upslope material and initiate a dry flow underneath. This kind of mechanisms could be at work today where new bright deposits are observed and may also contribute to some erosion such as in Terra Sirenum #3 and #6.

# 4  Summary and conclusion

We have performed an exhaustive survey of available near-infrared CRISM and OMEGA observations obtained at the 38 previously reported active gully sites located between latitudes 29° and 50° in the Southern Hemisphere. Information about the presence, absence and composition of ice where gullies are active is derived. The time sampling reaches 10° of solar longitude for several sites. An ice probability level that account for observational biases and limitations is associated with each change. These ice observational constraints are then compared to activity type.

Observations indicate the presence of either $CO_2$ or water ice for most sites and changes, with the exception of three minor activities. $CO_2$ ice is detected down to 32.3°S, equatorward of previous reports. All major erosive activities (formation of new channel, channel lengthening or widening, alcove widening) occur where (and, if applicable, when) $CO_2$ ice is observed or strongly probable. This result supports the idea that current channel development on Mars is mainly driven by $CO_2$ ice, as suggested by recent studies (Dundas, et al., 2015). At low latitudes (< 40°S), $CO_2$ ice thicknesses are estimated to be only 1 to 10 mm, which may be incompatible with some of the proposed $CO_2$-based formation mechanisms. Water ice is also present either simultaneously or alone prior and after $CO_2$ ice formation and could thus contribute to erosion.

We observe that $CO_2$ ice is weakly probable where and/or when some lower amplitude activity occurred. In particular, all new bright deposits are poorly compatible with $CO_2$ ice while they are linked with water ice. These bright deposits are preferentially observed on warmer nonpole-facing orientations and, at Gasa crater, in late winter around sublimation time. A specific formation mechanism linked with water ice sublimation could thus be acting at these locations.

Overall, our observations are consistent with the existence of several mechanisms currently participating to gully activity.

# 5  Acknowledgments

The author deeply thanks the CRISM and OMEGA scientific and engineering teams who make this work possible. All data used in this study are freely available through the Planetary Data System (http://pds-geosciences.wustl.edu/). The author would like to thank J. Carter, M. Cudel, C. Dundas, O. Harper, F. Morgan, S. Murchie, C. Pilorget and G. Poulleau for their specific help on this project. We finally thank C. Dundas, J. Dickson and an anonymous reviewer for their attentive reading of the manuscript and their helpful comments.